\documentclass[iop]{emulateapj}

\usepackage{multirow}
\usepackage{natbib}
\usepackage{longtable,rotate}

\newcommand{\noprint}[1]{}
\newcommand{\figsetstart}{{\bf Fig. Set} }
\newcommand{\figsetend}{}
\newcommand{\figsetgrpstart}{}
\newcommand{\figsetgrpend}{}
\newcommand{\figsetnum}[1]{{\bf #1.}}
\newcommand{\figsettitle}[1]{ {\bf #1} }
\newcommand{\figsetgrpnum}[1]{\noprint{#1}}
\newcommand{\figsetgrptitle}[1]{\noprint{#1}}
\newcommand{\figsetplot}[1]{\noprint{#1}}
\newcommand{\figsetgrpnote}[1]{\noprint{#1}}

\shorttitle{}
\shortauthors{M. S. Shaw et al} 
		
\slugcomment{Accepted for publication in The Astrophysical Journal}

\usepackage{lscape}

\begin{document}
\title{Spectroscopy of The Largest Ever $\gamma$-ray Selected BL Lac Sample}
\setlength{\paperheight}{11.0in}
\setlength{\paperwidth}{8.5in}

\author{Michael S. Shaw\altaffilmark{1}, Roger W. Romani\altaffilmark{1}, Garret Cotter\altaffilmark{2}, Stephen E.\ Healey\altaffilmark{1}, 
Peter F.\ Michelson\altaffilmark{1}, 
Anthony C.\ S.\ Readhead\altaffilmark{3}, Joseph L.\ Richards\altaffilmark{3}, 
Walter Max-Moerbeck\altaffilmark{3}, Oliver G.\ King\altaffilmark{3}, William J. Potter\altaffilmark{2}
}

\altaffiltext{1}{Department of Physics/KIPAC, Stanford University, Stanford, CA 94305}
\altaffiltext{2}{Department of Astrophysics, University of Oxford, Oxford OX1 3RH, UK}
\altaffiltext{3}{Department of Astronomy, California Institute of Technology, Pasadena, CA 91125}


\begin{abstract}
	We report on spectroscopic observations covering most of 
the 475 BL Lacs in the 2$^{nd}$ {\it Fermi} LAT catalog of AGN. Including
archival measurements (correcting several erroneous literature
values) we now have spectroscopic redshifts for 44\% of the BL Lacs. 
We establish firm lower redshift limits via intervening absorption systems and
statistical lower limits via searches for host galaxies for an additional
51\% of the sample leaving only 5\% of the BL Lacs unconstrained. The new
redshifts raise the median spectroscopic $\tilde z$ from 0.23 to 0.33 and 
include redshifts as large as $z=2.471$. Spectroscopic redshift minima from 
intervening absorbers have $\tilde z= 0.70$, showing a substantial fraction at large
$z$ and arguing against strong negative evolution. We find that detected BL
Lac hosts are bright ellipticals with black hole masses $M_\bullet$
$\sim 10^{8.5-9}$, substantially larger than the mean of optical AGN 
and LAT Flat Spectrum Radio Quasar samples. A slow increase in $M_\bullet$
with $z$ may be due to selection bias. We find that the power-law
dominance of the optical spectrum extends to extreme values, but this
does not strongly correlate with the $\gamma$-ray properties, suggesting
that strong beaming is the primary cause of the range in continuum dominance.
\end{abstract}



\section{Introduction}

The {\it Fermi} Second Source Catalog \citep[2FGL]{2FGL} lists the 1873 
most significant sources detected by the Large Area Telescope \citep[LAT]{atw09}
during {\it Fermi}'s first two years of sky survey observations. 
The majority of these sources are associated with jet-dominated Active Galactic
Nuclei, the so-called blazars, many of which are bright, compact radio sources.
There are, in fact, 1121 such associations (1017 at $|b|>10^\circ$), 
collected in the Second Catalog of 
AGN Detected by the {\it Fermi} LAT \citep[2LAC]{2LAC}. These AGN are further classified
as Flat-Spectrum Radio Quasars (FSRQ) where the optical spectrum is dominated
by thermal disk and broad-line region emission, BL Lacs (BLL), where the optical
spectrum is dominated by continuum synchrotron radiation, and a collection
of miscellaneous, mostly low luminosity related sources.  In 2LAC,
the sample included 410 BLL, 357 FSRQ, 28 AGN of other type (generally low $z$, 
lower luminosity Seyferts), and 326 AGN of (then) unknown type.  

	These `Blazars' (BLL and FSRQ) are the brightest extra-Galactic
point sources in the microwave and $\gamma$-ray bands; study of their
population and evolution are central topics in high energy astrophysics.
To support such studies we have acquired sensitive spectroscopic observations
of this sample.  In a companion paper \citep[][hereafter S12]{fsrq}, 
we reported on measurements of a large fraction of the FSRQ. Here we concentrate on
the BL Lac objects. Our study has also found types for some of 
the unclassified blazars; the `unknowns' have now decreased to 215 (19\%),
and the confirmed BLLs have increased to 475 (42\% of all 2LAC AGN).

In \S \ref{sec:obs}, we outline the sample properties, data collection, 
and data reduction steps. We also summarize principal features of 
the spectra. In \S \ref{sec:redshifts}, we describe our spectroscopic
constraints on the redshift, including a technique to provide uniform
redshift limits based on searches for host galaxy emission.
In \S \ref{sec:bh}, we give estimates of the BLL black hole masses. 
We turn to comments on the principal BLL feature, the non-thermal 
dominance in the optical in \S \ref{sec:ntd}, and conclude with general 
remarks in \S \ref{sec:conclusion}.

In this paper, we assume an approximate concordance cosmology -- 
$\Omega_m=0.3$, $\Omega_\Lambda=0.7$, and $H_0=70\ $km s$^{-1}$ Mpc$^{-1}$.


\section{Observations and Data Reduction}
\label{sec:obs}

\subsection{The BLL Sample}
\label{sec:sample}

BLLs were originally identified as optical violently variable
AGN, and are often characterized by an optical continuum dominated by
synchrotron emission. Their broad-band spectral energy distribution 
(SED) is described by a synchrotron component peaking in the far-IR to X-ray bands and 
an Inverse Compton (IC) component peaking in the MeV to TeV range.
In the radio these sources display strong core dominance. 
According to the unified model \citep{urr95} BLLs are the beamed counterparts 
of the FR I radio galaxy population, while the 
FSRQ are associated with FR IIs. However,
the principal BLL characteristic, a dominant and varying synchrotron/IC
continuum, is a sign of a powerful jet whose emission is beamed closely
toward the Earth line of sight. Thus the distinction between the
traditional BLL and the FSRQ is sensitive to the precise state
and orientation of the jet \citep[e.g.][]{giommi12} and, indeed, variations in 
jet power or direction can bring individual sources in or out of the BLL class (S12).

	Our own assignment of the BLL label follows
a pragmatic ``optical spectroscopic'' definition \citep{mar96}:
these are blazars with no emission lines greater than $5$ \AA\ observed equivalent width, 
and a limit on any possible 4000 \AA\ spectral break of $<40\%$ \citep{mar96, cgrabs,sha09}.
For transition objects with varying continuum emission, we retain the BLL label
if it has ever been confirmed to be a BLL with a high quality spectrum, even
if subsequently observed in an FSRQ state. Within the BLL population it is common to
classify sources based on the peak frequency of their SED's synchrotron component, as 
estimated from radio/optical/X-ray flux ratios, separating the sources 
into low-peak ($\nu_{peak} < 10^{14}$\,Hz, LBL), intermediate peak 
($10^{14} $\,Hz$< \nu_{peak}<10^{15}$\,Hz IBL), and high-peak ($\nu_{peak} > 10^{15}$\,Hz, HBL) 
sources.  We adopt here the LBL/IBL/HBL designations from 2LAC, which provides such subclasses for 74\% of
all 2LAC BLL, and 83\% of those with spectroscopic redshifts (Table \ref{tab:completeness}).

	The evolution of BLL has long been controversial, and it has been
claimed that they are predominantly a low redshift population, showing strong
`negative' evolution \citep[e.g.][]{bllev}, especially for the HBL class
\citep{rec00}. One challenge to studies of the cosmological evolution of these sources
is the difficulty of obtaining redshifts from their nearly featureless, continuum
dominated spectra. Indeed, many of the early studies using X-ray or radio-selected samples
had highly incomplete redshift measurements, even though the samples were confined to
relatively bright sources.  Uncertainty in
extrapolating from the measured set of redshifts complicated population interpretations. 
In the {\it Fermi} era, this issue becomes critical, as the large BLL contribution to
the LAT source population and the hard BLL $\gamma$-ray spectra ensure that these 
sources are a major fraction of the cosmic $\gamma$-ray background and may, indeed
dominate the LAT background at high energies (Ajello et al, in prep).

	Since the LAT provides a large, uniform, sensitivity limited 
($\sim$ flux limited) blazar sample, it provides a new opportunity to make progress
on these questions. Important to interpreting the LAT blazars are the strong
correlations between the LAT-detected (IC) part of the SED and the synchrotron
component covering the optical band. The synchrotron peak location determines
the sub-classification, but also correlates with the intensity and
LAT-band spectral index of the IC component, which in turn affects the depth 
of the LAT sample. Further, as illustrated in this paper, the synchrotron peak and
intensity also affect the difficulty of optical spectroscopic measurements. Since
we wish to recover the detailed evolution of the blazars, preferably also following 
differences among the LBL/IBL/HBL subclasses (or even better, a physical parent property
leading to these subclasses), one needs a detailed treatment of
both the $\gamma$-ray \citep[e.g.][]{aje12b}
and optical selection effects. We reserve such analysis for a future study, noting
here only the most prominent trends in the measured optical sample. Of course, characterization
and minimization of the optical biases are greatly aided by high redshift completeness,
the goal of the present paper.

	We have accordingly studied the 2LAC sample over a number of
years with a wide variety of telescopes, striving to be as complete as possible.
We here report on observations of 278 BLL objects and 19 other {\it Fermi} blazars, 
not included in S12. We further analyzed 75 SDSS spectra, treating them in the same 
manner as our new observations.  Note that, unless we suspected a published redshift 
was erroneous, we generally did not obtain new spectra of many of the brighter, famous BLL
with redshifts in the literature. In several cases when we obtained new data it strongly 
contradicted the literature redshift, either from a new secure spectroscopic
$z$ or from an intervening absorber at larger $z$. In the end we retained 107
redshifts from literature values; however for several early BLL $z$'s we have only 
inspected plotted spectra (of varying qualities); we suspect that at least a few 
erroneous values remain in this set. Our new data provide new spectroscopic
redshifts for 102 objects and secure lower limits for many more, as
summarized in Table \ref{tab:completeness}. This brings the 2LAC BLL sample to 
44\% redshift completeness, and 95\% completeness including redshift limits.

\begin{deluxetable}{lllll}
\tabletypesize{\small}
\tablecaption{Sample Completeness}
\tablecolumns{5}
\tablehead{
\colhead{Set} & \colhead{Total} & \colhead{Spec $z$}& \colhead{$z_{min}$} & \colhead{Unk. $z$}
}
\startdata
2LAC & 1121 & & & \\
BLL& ~\,475 &  209 & 241 & 25\\
\, LBL& ~\,~\,72 ({\it 21\%})  & ~\,35 ({\it 20\%}) & ~\,36 ({\it 23\%}) &  \\
\, IBL& ~\,~\,91 ({\it 26\%})  & ~\,41 ({\it 24\%})& ~\,49 ({\it 31\%}) &  \\
\, HBL& ~\,187 ({\it 53\%})    & ~\,98 ({\it 56\%}) & ~\,73 ({\it 46\%}) &  
\enddata
\tablecomments{BLL includes 65 AGN so classified since the 2LAC paper. 
349 BLL have 2LAC SED subclasses; percentages give the breakdown.
174 of the spectroscopic redshifts and 158 of the lower limits
have subclasses; percentage breakdowns are given. 
}
\label{tab:completeness}
\end{deluxetable}


\subsection{Observations}

	The quest for high completeness has driven us to employ medium and large 
telescopes in both hemispheres.  Observations were obtained with the Marcario Low 
Resolution Spectrograph (LRS) on the Hobby-Eberly Telescope (HET), with the ESO Faint 
Object Spectrograph and Camera \citep[EFOSC2]{buz84} and ESO Multi-Mode 
Instrument \citep[EMMI]{dek86} on the New Technology Telescope at La Silla 
Observatory (NTT), with the Goodman High Throughput Spectrograph (GHTS) on the 
Southern Astrophysical Research (SOAR) Telescope, with the Double Spectrograph 
(DBSP) on the 200'' Hale 
Telescope at Mt. Palomar, with the FOcal Reducer and low dispersion Spectrograph 
\citep[FORS2]{app98} on the Very Large Telescope at Paranal Observatory (VLT),
and with the Low Resolution Imaging Spectrograph (LRIS) at the W. M. Keck 
Observatory (WMKO). Observational configurations and objects observed are 
listed in Table \ref{tab:obs}.

	Since at the time of observation, many of these sources were not classified,
we often initially obtained only sufficient S/N to identify the redshift
of an FSRQ, or to firmly establish the source as a BLL. Also, with the variety of 
telescope configurations and varying observing conditions, 
the quality of the spectra are not uniform: resolutions vary from $4$ to $17$\,\AA\,,
exposure times from $180$\,s to $2400$\,s, and telescope diameters from 
$3.58$\,m to $10$\,m. S/N per resolution element varies from 10 to $>\ $300.
In a number of cases follow-on observations with higher S/N and/or 
higher spectral resolution allowed
us to more carefully study confirmed BLL lacking redshifts. Here we discuss
the most constraining spectrum or spectrum average for each source, referring
to this as the `primary' spectrum.

All spectra are taken at the parallactic angle, except for LRIS spectra using 
the atmospheric dispersion corrector, where we observed in a north-south 
configuration. In a few cases, we rotated the slit angle to minimize 
contamination from a nearby star. At least two exposures are taken of every 
target for cosmic ray cleaning. Typical exposure times are $2$x$900$\,s.

\begin{deluxetable*}{cccccccc}
\tabletypesize{\small}
\tablecaption{Observing Configurations}
\tablehead{
\colhead{Telescope} & \colhead{Instrument} & \colhead{Resolution} & \colhead{Slit Width} & \colhead{Objects} & \colhead{Filter} & \colhead{$\lambda_{min}$} & \colhead{$\lambda_{max}$} \\
\colhead{} & \colhead{} & \colhead{\AA} & \colhead{Arcseconds} & \colhead{} & \colhead{} & \colhead{\AA} & \colhead{\AA}
}
\startdata
HET & LRS & 15 & 2 & 41 & GG385 & 4150 & 10500\\
HET & LRS & 8 & 1 & 8 & GG385 & 4150 & 10500 \\
NTT & EFOSC2 & 16 & 1 & 31 & - & 3400 & 7400 \\
NTT & EMMI & 12 & 1 & 1 & - & 4000 & 9300 \\
Palomar 200'' & DBSP & 5 / 15 & 1 & 4 & - & 3100 & 8100\\
Palomar 200'' & DBSP & 5 / 15 & 1.5 & 5 & - & 3100 & 8100\\
Palomar 200'' & DBSP & 5 / 9 & 1.5 & 42 & - & 3100 & 8100\\
SOAR & GHTS & 6 & 0.84 & 2 & - & 3200 & 7200\\
VLT & FORS2 & 11 & 1 & 14 & - & 3400 & 9600 \\
VLT & FORS2 & 17 & 1.6 & 16 & - & 3400 & 9600\\
WMKO & LRIS & 4 / 7 & 1 & 90 & - & 3100 & 10500 \\
WMKO & LRIS & 4 / 9 & 1 & 40 & - & 3100 & 10500
\enddata
\tablecomments{For DBSP and LRIS the blue and red channels are split by a 
dichroic at 5600 \AA; the listed resolutions are for blue and red side,
respectively.}
\label{tab:obs}
\end{deluxetable*}

\subsection{Data Reduction Pipeline}
\label{sec:pipeline}

Data reduction was performed with the IRAF package \citep{tod86, val86} using 
standard techniques. Data was overscan (where applicable) and bias 
subtracted. Dome flats were taken at the beginning of every night, the spectral 
response was removed, and all data frames were flat-fielded.  Wavelength 
calibration employed arc lamp spectra and was confirmed with checks of night sky
lines. We employed an optimal extraction algorithm \citep{val92} to maximize 
the final signal to noise. For HET spectra, care was taken to use sky windows 
very near the longslit target position so as to minimize spectroscopic 
residuals caused by fringing in the red, whose removal is precluded by the
rapidly varying HET pupil. Spectra were visually cleaned of residual cosmic 
ray contamination affecting only individual exposures.  

We performed spectrophotometric calibration using standard stars from 
\citet{oke90} and \citet{boh07}. In most cases standard exposures were available
from the data night. On the queue-scheduled HET, and during our queue-scheduled 
VLT observations, standards from subsequent nights were sometimes used. 
At all other telescopes, multiple 
standard stars were observed per night under varying atmospheric conditions and
different air-masses. The sensitivity function was interpolated between standard 
star observations when the solution was found to vary significantly with time.

For blue objects, broad-coverage spectrographs can suffer significant second 
order contamination. In particular, the standard HET configuration using a
Schott GG385 long-pass filter permitted second-order effects redward of 7700 \AA.
The effect on object spectra were small, but for blue WD spectrophotometric
standards, second order corrections were needed for accurate determination of
the sensitivity function. This correction term was constructed following
\citet{fors}. In addition, since BLL spectra are generally simple power 
laws, we used our objects to monitor second order contamination and residual 
errors in the sensitivity function. In extreme cases, we fit an average deviation 
from power law across all objects in a night, and treated it as a correction 
to our spectrophotometric calibrations. This resulted in excellent, stable 
response functions for the major data sets.

Spectra were corrected for atmospheric extinction using standard values. We 
followed \citet{kri87} for WMKO LRIS spectra, and used the mean KPNO extinction 
table from IRAF for P200 DBSP spectra. Our NTT, VLT, SOAR, and HET spectra do not
extend into the UV and so suffer only minor atmospheric extinction. These
spectra were also corrected using the KPNO extinction tables.
We removed Galactic extinction using IRAF's de-reddening function and the Schlegel 
maps \citep{sch98}. We made no attempt to remove intrinsic reddening (i.e.: from 
the host galaxy).

Telluric templates were generated from the standard star observations in each 
night, with separate templates for the oxygen and water line complexes. We 
corrected separately for the telluric absorptions of these two species. We found
that most telluric features divided out well, with significant residuals only
apparent in spectra with high S/N. On the HET spectra, residual
second order contamination prevented complete removal of the strong water band 
red-ward of $9000$ \AA.

When we had multiple epochs of these final cleaned, flux-calibrated 
spectra with the same instrumental configuration, we checked for strong
continuum variation. Spectra with comparable fluxes were then combined
into a single best spectrum, with individual epochs weighted by S/N.

Due to variable slit losses and changing conditions between object and 
standard star exposures, we estimated that the accuracy of our absolute 
spectrophotometry is $\sim 30\%$ \citep{cgrabs}, although the relative 
spectrophotometry is considerably better.

\figsetstart
\figsetnum{1}
\figsettitle{Spectra}

\figsetend

\begin{figure*}
\epsscale{1.3}
\vspace{-30pt}
\hspace*{-40pt}
\plotone{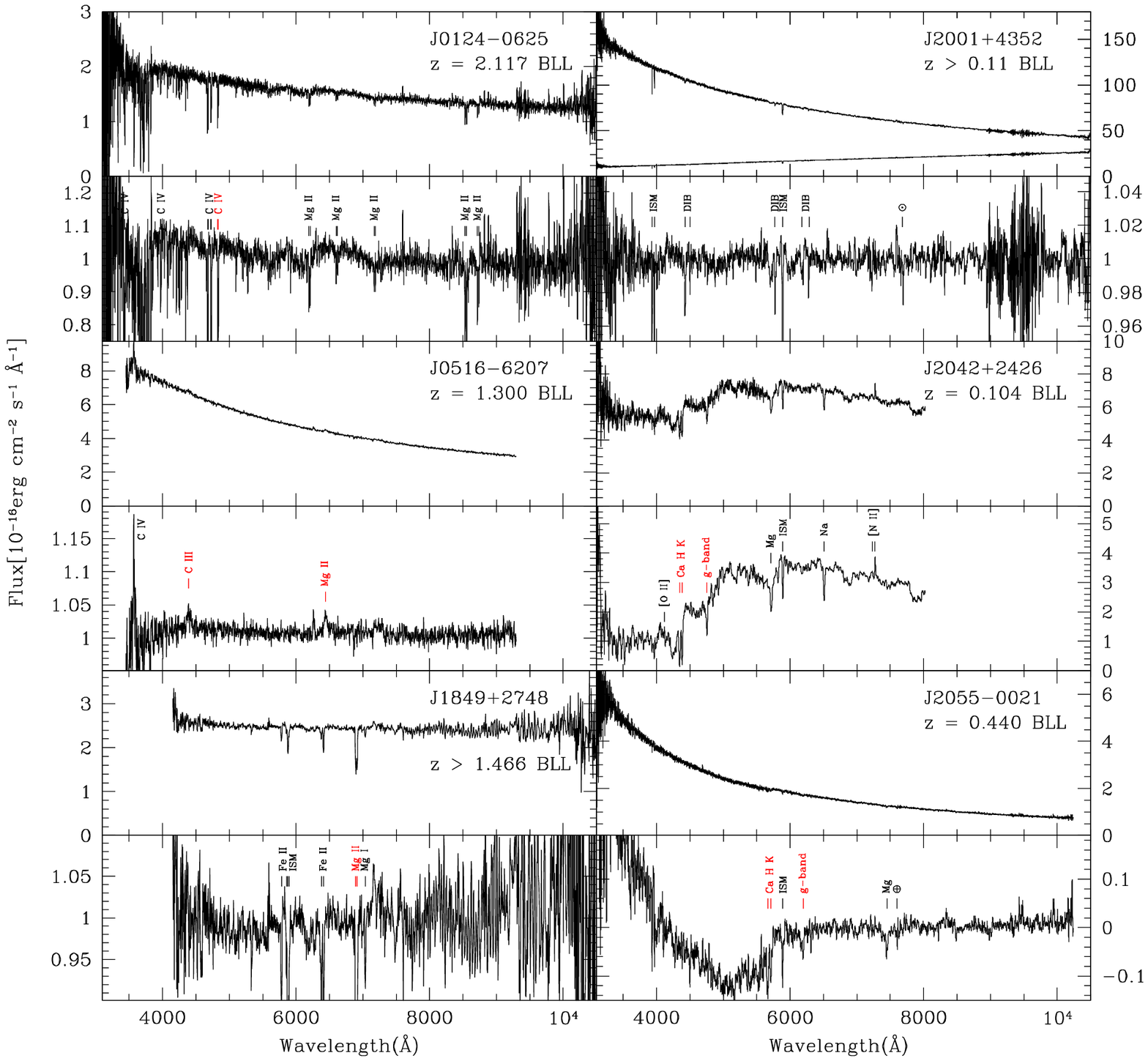}
\vspace{-40pt}
\caption{Spectra of {\it Fermi} BLLs.
Each object is presented twice  -- directly in the upper panel, and then with 
the best-fit power law `removed' (generally by division, but for composite
spectra by subtraction). This residual is plotted in the lower panel. SDSS 
spectra, while discussed in Table \ref{tab:props}, are not replotted in this Figure. The six
sample spectra here illustrate major trends. Similar figures for the full BLL
sample are available in the electronic edition of this journal. There
we mark only the two lines used to secure $z_{\rm spec}$; here other lines
of interest are marked and the two qualifying lines are marked in red. $z$
limits given are the most constraining limits presented---2 digits are given for limits
from non-detection of host galaxies; 3 digits, for intervening absorption systems. The
marked absorption system is the best spectroscopic limit; in some cases, a stronger
host galaxy limit is presented. See Table \ref{tab:props} for the precise spectroscopic limit.}
\label{fig:interesting}
\end{figure*}

\subsection{General Trends}

	To illustrate the principal trends in the BLL spectra we refer the reader to
Figure \ref{fig:interesting}. By definition, the dominant component is a power-law.
J0516$-$6207, however, shows that after removal of a power-law weak, but broad
 \ion{C}{4}, \ion{C}{3}, and \ion{Mg}{2} features may occasionally be seen in 
high S/N spectra. Here the equivalent widths ($<1$ \AA) are sufficiently small 
to secure the identification as a BLL. However, should the continuum fade by 
$\sim 10\times$, this would be classified (at that epoch) as an FSRQ. Such
transition objects support the idea of a Blazar continuum, rather than two
distinct populations \citep{fos98,ghi08}. In S12, we reported significant 
broad line measurements for 5 of our BLL, including J0516-6207.

	While the power-laws of most BLL are very blue, like J0516$-$6207, a few
like J1849+2748, appear intrinsically flat or red, even after correcting for Galactic 
extinction. This may plausibly be a sign of a synchrotron component peaking near the
optical, but might also indicate incomplete extinction correction, with residual reddening
caused by dust not in the \citet{sch98} maps. It could also be intrinsic host extinction.

	The Galactic reddening can be very severe. For J2001+4352 (upper right) we show
both the highly extincted, pre-correction spectrum and the blue post-correction power law.
This source is in a direction of known high $A_V=1.75$. Such extincted power-law spectra provide
an excellent opportunity for ISM studies:  The features seen after de-extinction and division
by the best-fit power law (lower panel) are all interstellar in origin -- Galactic H and K, 
\ion{Na}{1} 5892, and a series of diffuse interstellar bands, as described in \citet{yua12}.

For J0124$-$0625 (upper left) the residual absorption features are intergalactic in origin.
Redward of $3900$ \AA\ we detect a number of metal-line systems, blueward one sees the onset
of strong Lyman-$\alpha$ forest absorptions. These features determine a redshift $z = 2.117$
(one of the highest in our BLL sample). The lack of similar Ly$\alpha$ forest absorption in
many of our other high S/N, high resolution spectra allows us to place statistical upper
limits on the redshift as described in \S \ref{sec:upper}.

Finally at lower right we see two BLL with significant flux from the host.  In J2042+2426, 
the galaxy provides about a third of the total flux and is easily visible in the raw spectrum.
This is still safely a BLL, and we can measure the continuum contribution by a 
`Non-Thermal Dominance'
(see \S \ref{sec:ntd}, here NTD=1.38). For J2055$-$0021, the host is swamped by the core 
synchrotron emission (NTD=47.5) and the galaxy features are visible only after 
subtracting the best-fit power law, as in the lower right panel.  The flux increase
in the blue appears due to residual few-\% fluxing errors (here, likely incomplete 
correction for atmospheric extinction), rather than intrinsic emission. Despite the careful calibration,
such residual fluxing issues persist in several spectra. However, the high-pass filtering 
described in \S \ref{sec:hostfit} ensures that our measurements of, and bounds on,
host galaxy flux are almost completely immune to such residual calibration artifacts.
We find that a number of BLL show visible host galaxy components, all consistent with 
giant ellipticals \citep{urr00}.  We discuss the flux distribution of these host 
galaxies in \S \ref{sec:lower}. 

\subsection{Individual Objects}
\label{sec:obs_individual}

A number of BLL reductions required special treatment. For example a 
few objects clearly required changes to the Schlegel map $A_V$, so that 
the de-extinction resulted in clean power laws.  For J0007+4712, $A_V$ 
was increased from 0.3 to 0.8 and for J1941$-$6211 from 0.3 to 1.0. 
Conversely we decreased the $A_V$ of J1603$-$4904 from 7.8 to 5.0
and J2025+3343, from 6.15 to 5.0. We checked the recent recalibration of the Galactic
extinction maps \citep{sf11}, but did not find large changes, so these extinction features
affecting our BLL are probably on scales below the map resolution.

J1330+7001 was observed off of the parallactic angle---the ensuing drop in blue flux is not intrinsic to the system, and our power law is fit only redward of $5000$ \AA. We thus remove the broadband residual
in our power law divided spectrum in Figure \ref{fig:interesting}. J1829+2729 and a nearby star were spatially unresolved in our data. The presented spectrum is a composite of starlight and quasar light, which the significant emission features all identified as $z=0$ stellar or ISM features.

In a few cases, objects previously cataloged as BLLs do not have sufficient S/N in our spectra
for a definitive BLL classification.  For J0801+4401, we find that undetected broad lines
could have an equivalent width as large as 9.5 \AA; for J0209-5229 the limit is 5.6 \AA, 
for J1311+0035 the limit is 8.0 \AA\ and for J1530+5736 we could have missed lines 
as strong as EW=5.5 \AA. As higher S/N spectroscopy would likely confirm the 
archival BLL designations, we consider them BLLs for the purposes of this paper.

Five of the BLL described here had high significance broad line detections and have
already been described in S12; we re-measure these spectra here for a uniform BLL treatment.
In J0847$-$2337 and J0430$-$2507, the flux and spectral index measurements differ from the
S12 values. This is because in the present analysis we first subtract the host galaxy flux, 
and calculate the flux and spectral index of the remaining non-thermal component. In 
S12, no such correction was attempted.

\section{Measuring BLL Redshifts}
\label{sec:redshifts}

	The opportunity to advance our understanding of BLL evolution with the large,
flux limited {\it Fermi} sample is important \citep{aje12}. Yet, despite the substantial
telescope resources and careful analysis summarized above, many BLL did not yield
direct spectroscopic redshifts, due to the extreme weakness of their emission 
lines \citep{sba05} and lack of clear host features.
Therefore we collect here both the direct redshift measurements and quantitative
constraints on the allowed redshift range for our observed BLL.

\subsection{Emission Line Redshifts}
\label{sec:emission}

We visually inspected all spectra for AGN emission line features,
and host galaxy absorptions. Spectroscopic redshifts are measured by cross-correlation analysis using the 
rvsao package \citep{rvsao}. We require one emission line to be present at the $>5\sigma$ level, and a second line present at the $>3\sigma$ level---significances are measured by fitting a Gaussian template in the splot tool; we allow the width and amplitude of the Gaussian to vary, but fix the center at the redshift derived by rvsao's xcsao routine. For this study, we limited our search to typically strong emission lines known to be present in some BL Lacs: Broad emission from \ion{C}{4} (1549, 1551 \AA), \ion{C}{3} (1909 \AA), \ion{Mg}{2} (2796, 2799, 2804 \AA), H$\gamma$ (4340 \AA), H$\beta$ (4861 \AA), and H$\alpha$ (6563 \AA) and narrow emission from [\ion{O}{2}] (3727, 3729 \AA), [\ion{O}{3}] (4959, 5007 \AA), and [\ion{N}{2}] (6549, 6583 \AA). While other species exist in our spectra, these here listed are sufficient for definite redshift IDs. Velocities are not corrected to helio-centric or LSR frames.

In many cases, spectroscopic redshifts are further determined by a significant ($>3\sigma$) detection of a host galaxy, as will be described in \S\ref{sec:hostfit}. A few redshifts require further note: In J0124-0625 and J1451+5201, we identify the redshift by a Ly$\alpha$ and \ion{C}{4} 
absorption system at the onset of the Ly$\alpha$ forest. 
In J0434-2015, we identify a 
single strong feature with [\ion{O}{2}], consistent with weak \ion{Mg}{2} and Ca H/K 
absorptions. For J1728+1215, we 
find strong \ion{Mg}{2}, confirmed by [\ion{O}{2}] at the same $z$ in an archival 
spectrum. In J2152+1734, we identify a strong feature with \ion{Mg}{2} confirmed by a 
significant [\ion{O}{2}] detection in archival spectroscopy.

For a few objects only a single emission line was measured with high S/N. In general
we use the lack of otherwise expected features to identify the species
and the redshift with high confidence. Nevertheless, a few redshifts have some 
systematic uncertainty and are marked by a `:' in Table \ref{tab:props}. We briefly 
discuss these cases here. For J0007+4712, we derive a redshift from the clear onset of 
the Lyman-$\alpha$ forest and report only two significant figures. In J0212+2244, we determine a 
tentative $z$ from weak Ca H, K and g-band absorptions.  For J0439-4522, we identify the one strong emission feature 
as \ion{C}{4}; intervening absorption excludes a \ion{Mg}{2} identification, but a 
less likely \ion{C}{3} identification at $z\sim 1.45$ is not conclusively ruled out. J0629$-$1959 presents broad but weak emission at the redshift of the 
highest $z$ metal line absorption system (1.724). We thus identify this, tentatively, as 
the object's true $z$.  For 
J0709$-$0255, we identify the strong feature at 9200 \AA\ with [\ion{O}{3}] by the 
line shape; an [\ion{O}{2}] identification at $z \sim 0.84$ is not excluded. For J0825+0309, we find
significant [\ion{O}{3}] emission at 5007 \AA\ (and possible, but not significant emission at 4959 \AA),
at a $z$ consistent with an \ion{Mg}{2} feature identified in \citet{sti93}. 
We find weak features in J1231+2847 at the SDSS $z$, but they have 
low significance.
For J1312$-$2156, we find a plausible \ion{Mg}{2} feature;
this single line identification is in a small allowed redshift range ($z \sim 1.6$),
other identifications for this line are spectroscopically excluded.  In J1754-6423, we
tentatively identify emission at $\sim 6300$ \AA\ with 
\ion{Mg}{2}---higher $z$ redshifts are excluded by the lack of Ly$\alpha$ forest. In J2116+3339's
spectrum, a significant broad emission feature is identified with \ion{C}{4}, consistent with a weak bump in the far blue
at Ly$\alpha$. Nevertheless a lower $z$ 
redshift is possible if the purported Ly$\alpha$ line is not real. J2208+6519 presents one strong, broad
emission feature, tentatively identified as \ion{Mg}{2}---a \ion{C}{4} identification at $z \sim 1.8$ is not
excluded.

\subsection{Intervening Absorbers}
\label{sec:intervening}

For some BLLs, the core light passes near an intervening galaxy on its way to Earth. 
At small radii one can encounter low excitation clouds in the galaxy's halo, giving absorption doublets 
from \ion{Mg}{2} at (2795.5, 2802.7) \AA . Larger impact parameters can sample 
\ion{C}{4} at (1548.2, 1550.77) \AA. In some low excitation (i.e.: \ion{Mg}{2}) systems, we also
see absorption from \ion{Fe}{2} at (2344.2, 2374.4, 2382.7, 2586.6, 2600.1) \AA. Finally, for 
our highest redshift BLL we Lyman-$\alpha$ absorption at 1215 \AA\, for the metal line systems,
as well as onset of the Lyman-$\alpha$ forest.

For unsaturated absorptions, the doublet ratio for \ion{Mg}{2} and \ion{C}{4} is 2:1, 
with the blue line dominant. In saturated absorptions, the ratio is 1:1 \citep{nes05,mic88}. 

We visually search all spectra for candidate doublets, and follow \citet{nes05} in employing 
a quantitative test of the significance of each candidate. We used a 
two Gaussian template with wavelength spacing scaling with $z$, but free amplitudes, and
fit the equivalent width and error of each component, using the splot tool in iraf.
For the candidate to qualify as a detection we require the stronger (bluer) line to 
have $>5\sigma$ significance, and the second line to have $>3\sigma$ significance. In
a few cases, one component of an otherwise strong doublet was affected by skylines,
telluric features or cosmic rays. In these cases, another expected feature from
the absorption complex (e.g.: a \ion{Fe}{2} line) detected at $>3\sigma$ qualified the system.
We further require the doublet ratio to be consistent (within errors) to a value 
between 2:1 and 1:1.

	Our principal goal is not an absorption line study. Thus we concentrated on the
longest wavelength (highest $z$) candidate system and measured sufficient lines to confirm
the $z$ (i.e: two significant lines). After validation we continued to search for
higher $z$ until no candidates passed the significance test.  We therefore believe that 
we have found the highest $z$ absorption system in each of our spectra strong enough to 
reach the $5\sigma/3\sigma$ criteria above.

	Since we see the onset of the Lyman-$\alpha$ forest in our highest redshift 
objects, the red end of the forest provides a strict lower limit on redshift.This 
can be higher than that inferred from the reddest metal line system.

	We list these spectroscopic minimum $z$'s as $z_{\rm min}$, when available,
in Table \ref{tab:props}.

\subsection{Redshift Upper Limits}
\label{sec:upper}

	We can use the absence of Lyman-$\alpha$ absorptions to provide
statistically-based upper limits on $z$ for all BLLs without redshift. 
The exclusion $z_{\rm max}$ depends on the spectral range, resolution and 
S/N of the particular observation, but is generally $1.65 < z < 3.0$. 

	To quantify the upper bound, we need the expected density of Ly$\alpha$ forest
absorbers. \citet{pen04} find that for rest EW $\ge 0.24$ \AA, 
$dN/dz \sim 40$ at $z = 1.6$, varying with redshift as $\log dN/dz \propto 1.85 \log(1+z)$. 
We follow \citet{wey98} for the EW scaling: $dN/dz \propto e^{-(EW_{\rm rest} - 0.24)/0.267}$. As 
the S/N and resolution of our data vary, we generally measured a conservative 
uniform $3\sigma$ sensitivity for narrow Ly$\alpha$ absorptions $100$ \AA\ from 
the blue end of our spectra. Typical EW limits in this range were $\sim 0.2-1.0$ \AA. 
Given this density we solve for the $\Delta z$ range giving a Poisson probability 
of $0.32$ (ie: $1\sigma$) for detecting no absorbers, obtaining:
\begin{equation}
\Delta z = 0.167 \cdot e^{(EW_{\rm rest} - 0.24) / 0.267)} \cdot \left(\frac{\lambda_{\rm min}}{1215}\right)^{-1.85}
\label{eqn:upper}
\end{equation}
where $\lambda_{\rm min}$ is the effective blue end of the spectrum 
(generally 3150 -- 4200 \AA), and EW is the measured equivalent width limit.
Thus, we infer a maximum source redshift $z_{\rm max}= (\lambda_{\rm min}/1215) - 1 + \Delta z$. 
In some cases, the S/N is too low at the blue end of the spectrum. We then 
measure EW limits closer to the sensitivity peak of the spectrum. Of course,
with a larger $\lambda_{\rm min}$ for the effective spectrum end
Equation \ref{eqn:upper} gives less constraining upper limits. 

        If the actual blazar redshift $z$ is very close
to $z_{\rm max}$ as estimated above, then its UV radiation may photo-ionize
Ly$\alpha$ clouds along the line of sight, postponing the onset of the forest
and artificially lowering $z_{max}$. In practice, the effect is usually small
($\Delta z_{max} \sim 0.01-0.02$) except for large $z_{max}$ when our bound is 
generally not of interest.  We follow \citet{bdo88}
to estimate this `proximity effect' correction, by computing
\begin{equation}
\omega(z_{\rm Ly\alpha})={{f_\nu} \over {4\pi J_\nu}} { {(1+z_{\rm Ly\alpha})^5}\over {(1+z)}}
\left [
{ {(1+z)^{1/2}-1} \over {(1+z)^{1/2}-(1+z_{\rm Ly\alpha})^{1/2}} }
\right ]^2
\label{eqn:prox}
\end{equation}
where $f_\nu$ is the blazar flux at the absorption Lyman limit (at $z_{\rm Ly\alpha}$) and
$J_\nu = 10^{-21.5} {\rm erg\,cm^{-2}\,s^{-1}\,Hz^{-1}\,sr^{-1}}$ is 
the cosmic ionizing flux, conservatively estimated for our redshift range.
We compute $f_\nu$ from the power law fit in Table \ref{tab:props}, and increase the
blazar redshift $z$ until $\omega(z_{\rm Ly\alpha})<1$. We thus adopt these revised 
$z_{max} = z$ and quote these corrected upper limits in Table \ref{tab:props}.

\subsection{Host Galaxy Fitting}
\label{sec:hostfit}

\begin{figure*}
\vspace{-10pt}
\hspace*{-30pt}
\epsscale{1.3}
\plotone{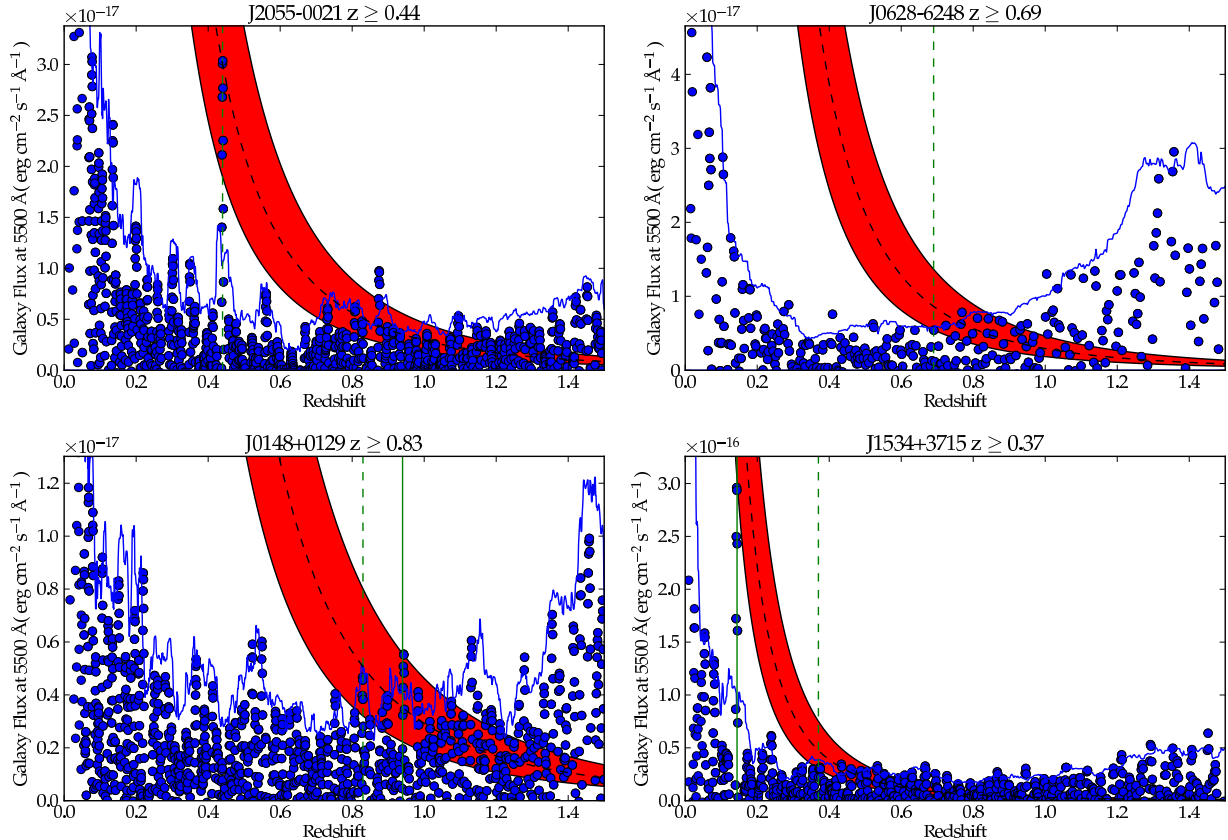}
\vspace{-30pt}
\caption{Best fit host galaxy fluxes as a function  of trial $z_i$ (blue dots). Our estimate 
of the $2\sigma$ local systematic flux error is shown by the blue line
(see text). The red bands give the $1\sigma$ range about the expected host flux 
for $M_R=-22.5 \pm 0.5$.  Vertical green dashed lines give our $z_{\rm min}(-22.5)$ 
limit on the redshift.  For comparison, solid green lines give the spectroscopic
redshifts, when measured. For J2055$-$0021 and J0148+0129 these are consistent; 
the former has a high significance host detection (Figure \ref{fig:interesting}), 
while for the latter the higher $z$ redshift is from emission lines.
J1534+3715 is one of 9 objects with limits inconsistent with spectroscopic redshifts. 
Here, the host galaxy is sub-luminous ($M_R = -21.87\pm0.16$) and thus (just) missed 
by this technique (Prob = 0.15).}
\label{fig:fitter}
\end{figure*}

It has been claimed that BL Lac objects are hosted by giant elliptical 
galaxies with bright absolute magnitude -- $M_R = -22.9 \pm 0.5$ in our 
cosmology \citep{urr00,sba05}. 

If we adopt the common assumption that these are standard candles, we 
can estimate the redshift of the BLL by detecting such galaxies. 
In imaging studies, one looks in the wings of the BLL for the host 
galaxy flux, and compares that to the standard candle flux at various 
redshifts \citep{sbaim,mei10}. In spectroscopic studies, one typically 
looks for individual absorption features (i.e.: H, K, g-band). One
can also use the lack of such lines as evidence that the BLL is 
at higher redshift \citep{sba05,sha09}.

	With high S/N spectra, however, one can obtain more stringent limits
by using the entire elliptical template rather than just one (or a few) 
lines. \citet{plo10} developed a technique of fitting for host galaxies in
SDSS BLL spectra. We expand here on that method for our more 
heterogeneous spectra.

Our spectra come from a variety of spectrographs in disparate observing 
conditions and we find low frequency systematic fluctuations in many of the fits. 
These are likely caused by imperfect spectrophotometric fluxing and second order 
contamination as discussed in \S \ref{sec:pipeline}. These effects can dominate 
over real galaxy features.  Using SciPy's Signal Processing 
routines\footnote{Documentation and more information available at 
http://docs.scipy.org/doc/scipy/reference/signal.html}, we construct a 
bandpass Kaiser window from 220\,\AA\, to $1.5\times$ the Nyquist frequency. We apply that 
window as an effective high pass filter both to our spectra and to the 
templates, to mitigate this low-frequency noise before fitting \citep{kai80}.

We test possible redshifts $z_i$ on a grid scaled to the spectrograph
resolution with constant spacing in $\log z$. This grid is thus
$z_i= (\frac{2\Delta\lambda}{\lambda_0}+1)^i - 1$, where $\Delta\lambda$ 
is the pixel scale of the spectrograph.  For each trial $z_i$ we fit 
the power law $F_\nu \propto \nu ^\alpha$ index and flux and the amplitude 
of a redshifted elliptical template.  This host template is generated from the
PEGASE model \citep{fio97} tables and evolved to low $z$ from $z=2$, as in \citet{odo05}.
For uniformity, we here use the same template for all $z_i$, and do not perform
any evolution corrections.

Our fit minimizes $\chi^2$ with three free variables at each trial $z_i$. We employ 
the scipy.optimize.leastsq routine based on a Levenberg-Marquardt fitter.
 
To model host slit losses, we assume an $r=10$\,kpc de Vaucouleurs
profile with Sersic index 1/4 \citep{odo05} and account for the individual observations' 
slit width and seeing profile (measured from the core full width at half maximum, FWHM). 
Since we have employed optimal extractions of the BLL spectra, our effective 
aperture along the slit varies, but we estimate a typical width of $\approx 2\times$ 
the spatial FHWM achieved during our spectral integration.
Accordingly we estimate host slit losses through a rectangular aperture of the
slit width $\times$ twice the spectrum FWHM. Inferred host fluxes are corrected
for these slit losses.

Results of sample fits are shown as the blue dots in Figure \ref{fig:fitter} where the 
fit amplitude of the host galaxy template is plotted against trial redshift. 

\subsection{Power Law Fit}

We report the power law fluxes and spectral indices of the best fit to the 
de-extincted spectrum in Table \ref{tab:props}. The flux is given in units of 
Log $10^{-28}$erg cm$^{-2}$s$^{-1}$Hz$^{-1}$ as observed at $10^{14.7}$\,Hz ($\sim 5980$ \AA), 
the center of our typical spectral range. The index $\alpha$ is measured $ F_\nu \propto
\nu^\alpha$. These values may be combined with multi-wavelength data to 
study the continuum SED of the blazars in our sample. Since the statistical errors
on the fit are, in general, unphysically small, we follow S12 in estimating
errors on the spectral index by independently fitting the red and blue halves
of the spectrum.  Note that large errors bars generally indicate 
a relatively poor fit to a simple power law rather than large statistical errors.
The statistical errors on the $F_\nu$ amplitude are also small; we convolve these
with our estimated $30\%$ overall fluxing uncertainty \citep{cgrabs}, which
dominates in nearly all cases.

\begin{figure}
\vspace{-0pt}
\hspace*{-0pt}
\epsscale{1.1}
\plotone{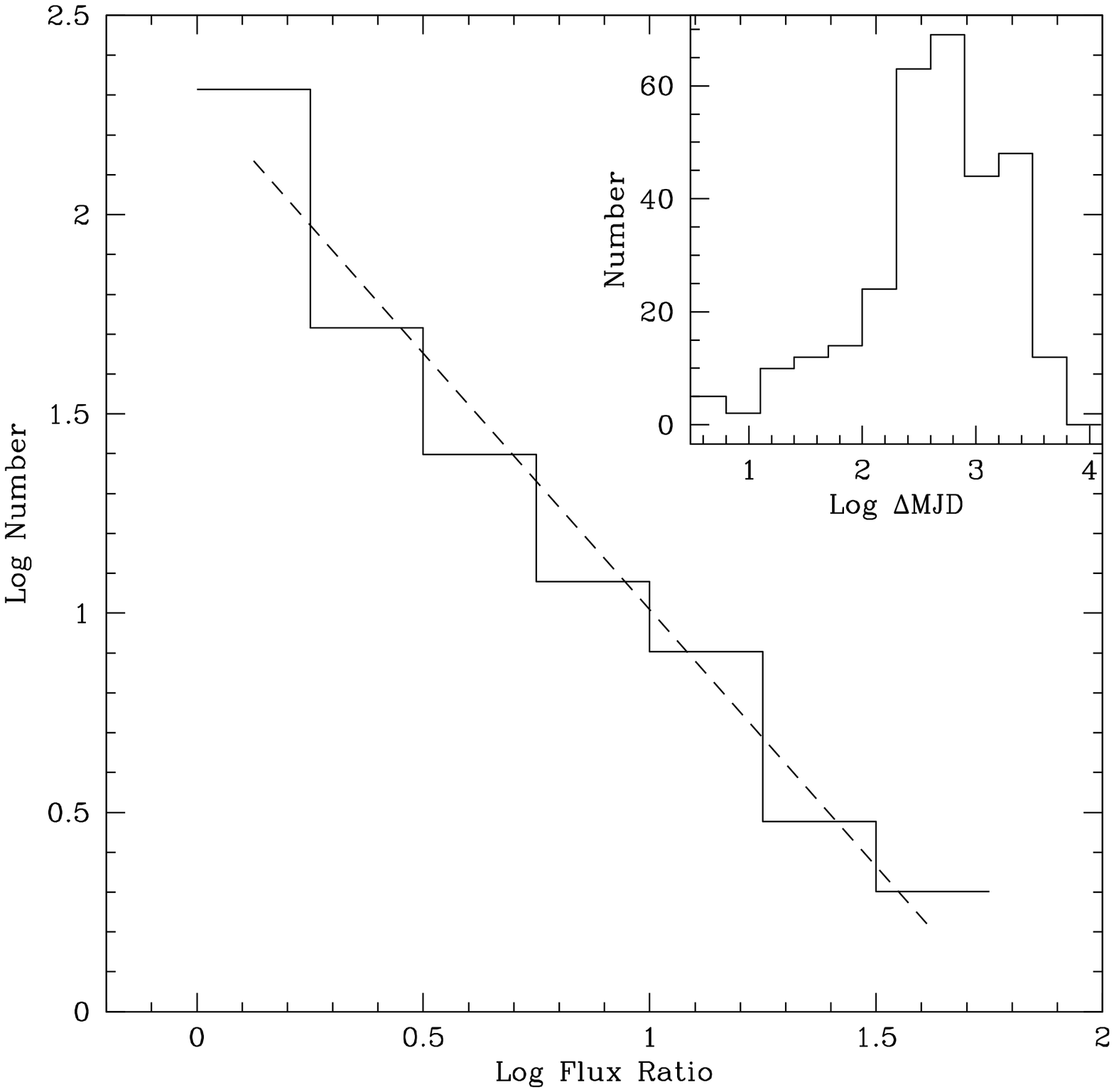}
\vspace{-0pt}
\caption{Histogram of maximum flux ratios. For each object where we have 
collected multiple spectra, we compute the flux ratio between the power law components 
of the brightest and faintest observed epoch. These are plotted as a log-log histogram.
The dashed line is the best-fit power law with index $\alpha=1.29$. The inset 
histogram shows the distribution of time between observations -- typically $\sim$ 1 year} 
\label{fig:variability}
\end{figure}

For objects with high significance ($>3\sigma$) detections of galaxies, as described 
in \S \ref{sec:lower}, we report the best fit power law from the simultaneous 
power-law/host fit in \S \ref{sec:hostfit}. 

When we have observed objects at multiple epochs, we also fit a power law to the 
other, non-primary spectra. These fluxes vary substantially, some by more than $10\times$. 
In Figure \ref{fig:variability}, we show the distribution of $f_{\rm max}/f_{\rm min}$
flux ratios. This is well-described by an $\alpha=1.29$ power law. Epochs from our fiducial spectra
are listed in Table 3.

\subsection{Testing the Standard Candle Assumption}
\label{sec:stdcandle}

\begin{figure}
\vspace{-0pt}
\hspace*{-0pt}
\epsscale{1.1}
\plotone{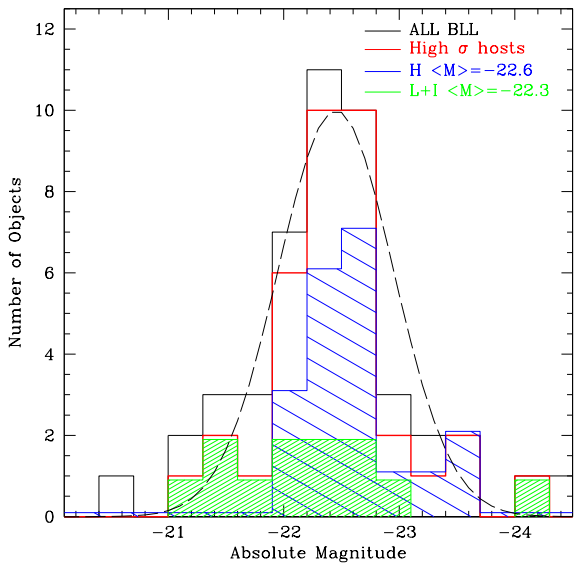}
\vspace{-0pt}
\caption{
Measured BLL host absolute magnitudes ($R$ equivalent at $z=0$).
Black histogram: all BLL hosts with spectroscopic redshift.
A weighted $\chi^2$ fit gives $M_R = -22.5 \pm 0.5$, our best estimate for 
the luminosity of BLL host galaxies.
Red: hosts with high significance ($>3\sigma$) detections. 
Green and Blue sub-histograms show high significance LBL+IBL and HBL hosts, 
respectively.
} 
\label{fig:candlemag}
\end{figure}

	BLL with a redshift and a secure ($>3\sigma$) host detection can be used
to test the uniformity of the host luminosities. There are 59 such BLL in our sample.
We derive synthetic $R$-band magnitudes by applying a Kron-Cousins R filter to 
our spectra \citep{mei10}. The results are shown as a histogram in 
Figure \ref{fig:candlemag}. We find $<M_R> = -22.5$, down $\sim 0.4$ magnitudes 
from $M_R = -22.9$ found in \citet{sba05}. We find a similar spread in luminosity 
($\sim \pm 0.5$ magnitudes). When we separate the host measurements for
lower-peak sources (LBL+IBL) we find that they have a median luminosity
$\sim 0.3$ magnitudes fainter than that of our HBL. Unfortunately, we do not have 
sufficient LBL+IBL hosts to test for such differences at high significance. 
Past studies differ: \citet{urr00} found no significant offset in the host 
magnitudes of HBL and LBL, but \citet{sba05} noted that higher peak HBL 
tend to have more luminous hosts.

Two of the high significance host galaxies have imaging magnitudes reported 
in \citet{sba05}. For J1442+1200, we measure $M_R = -22.99 \pm 0.14$; 
\citet{sba05} reported $M_R = -22.77$. For J1428+4240, we measure $M_R = -22.78 \pm 0.13$; 
\citet{sba05} find $M_R = -22.75$. These are consistent within measurement errors,
a good check of our slit-loss corrections and magnitude estimates.

Overall, the LAT BLL sample thus represents a fainter host population of than those
studied in previous work. Conceivably a higher (LBL+IBL)/HBL ratio
in our sample causes part of the difference (although we remain HBL dominated). However,
we suspect that our rather exhaustive $8$-m class campaign, skipping most objects 
with redshifts already in the literature, selects for fainter host galaxies 
than in the past. Thus we may be probing fainter on the true host luminosity
distribution;  the BLL for which we were not able to provide host detections 
may then be similarly under-luminous compared to previous studies. A true
evaluation of the intrinsic host luminosity distribution, as well as any
dependence on subclass type, will require a careful assessment of the
parent population ($\gamma$-ray) and host detection selection biases.

In the rest of this section, we conservatively adopt our $M_R = -22.5 \pm 0.5$ estimate. 
We do also report (Table \ref{tab:props}) more aggressive redshift limits based on the common 
assumption $M_R = -22.9 \pm 0.5$, for more direct comparison to previous work,
but we recommend adoption of the less stringent  $M_R = -22.5 \pm 0.5$ redshift 
constraints.

\subsection{Lower Limits from Non-detections of Host Galaxy}
\label{sec:lower}

We use the results of the fitting in \S \ref{sec:hostfit} and the calibration 
in \S \ref{sec:stdcandle} to constrain the redshift of the host. At each trial
redshift, our fitter reports a flux ($f \pm \sigma f$) for the host galaxy. This
is to be compared with the model flux from the
redshifted standard candle elliptical template ($f_M \pm \Delta f_M$).

	While we have greatly decreased the effect of the low frequency noise 
in our spectral fits using the high-pass filter, we still find that the	
statistical errors on the fit host galaxy fluxes are unrealistically small;
these flux estimates remain dominated by systematic effects.
We therefore construct an error estimate $\Delta f$ for each $z_i$ by measuring the
dispersion of flux estimates for nearby redshift bins. This is computed
from a sample of the 30 nearest $f(z)$, skipping 5 bins on each side of
our test $z_i$ to minimize high pass correlation. After $\sigma$-clipping,
the fit flux distributions are well behaved and we use these to compute a 
2-$\sigma$ upper limit (scaled from the rms) at each $z_i$. Vectors
of these upper limits are shown by the jagged blue lines in Figure \ref{fig:fitter}.
These have captured the local effective noise quite well and so we adopt
these vectors as effective $2\sigma$ confidence limits.

Fit fluxes well above these $2\sigma$ limits denote likely host detections.
Indeed, we found that this automatic processing was quite effective at
flagging candidate $z_i$ for host detection. Here, however, we focus on
how well a fit flux $f$ with local effective error $\Delta f$ can be used to
{\it exclude} a host of the expected magnitude $f_M$ at the test redshift $z$. Thus
we compute a probability that the fit flux $f$ and error $\Delta f$ are consistent
with the expected model flux $f_M$ and its uncertainty at the given $z$ as
\begin{equation}
Prob(z; f, \Delta f) = C \int_{f_M}^\infty G(f^\prime_M, \Delta f^\prime_M ) \cdot G(f, \Delta f) df'_M
\end{equation}
where $G(x,\sigma_x)$ is a Gaussian of width $\sigma_x$ centered at $x$.
The normalization $C$ is chosen such that 
$Prob = 1$ for $f_M = 0$ (i.e.: any $f$ is acceptable for a model of zero flux). 
This is a conservative choice as it does not exclude over-luminous hosts.
For example, when we assume a model $M_R=-22.5 \pm 0.5$ this probability
also finds any $f$ consistent with $M_R = -22.9 \pm 0.5$ to be acceptable.
When the fitter returns an unphysical negative $f$, we evaluate the probability
for $f=0$ and the local $\Delta f$.  

	This probability becomes substantial for $z$ near a good candidate redshift.
It also grows as the sensitivity of our host search drops at large $z$.
We thus calculate a minimum redshift ($z_{min(-22.5)}$) 
as the lowest redshift for which $Prob \ge 0.17$ (i.e.: $1\sigma$). We list 
these values in Table \ref{tab:props}. For comparison, we also give $z_{min(-22.9)}$ 
calculated in the same fashion, assuming a model $M_R = -22.9 \pm 0.5$. 
Comparison between the spectroscopic detections and $z_{\rm min}$ suggest
that the $M_R = -22.5$ value is most consistent with observed detections, and
lower bounds (as expected from \S3.6). We recommend use of these conservative lower limits.
Note also that the vector $Prob(z)$, once normalized with an appropriate prior and truncated
at $z_{\rm max}$, can be used as a PDF for the BLL redshift. 

\subsection{Redshift Distribution}

\begin{figure}
\vspace{-0pt}
\hspace*{-3pt}
\epsscale{1.2}
\plotone{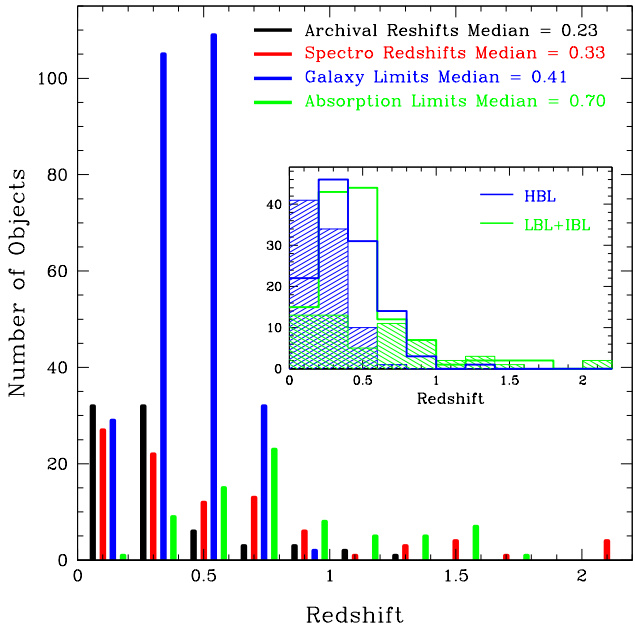}
\vspace{-0pt}
\caption{BLL redshift and redshift lower limit distributions. Within each $\delta z=0.2$ bin,
we show (left to right) the Archival $z$, our new spectroscopic $z$, lower limit $z$'s from
host fitting and lower limit $z$'s  from intervening absorption line systems. The limits show that
the spectroscopic redshift samples, particularly the archival sample, are selection biased
to low $z$.
The subframe shows the redshifts (filled histograms) and lower limits (open histograms)
for the LBL+IBL (green) and HBL (blue). Again, both limit histograms extend to higher $z$.
The LBL+IBL sample extends somewhat further than the HBL.
}
\label{fig:z_hist}
\end{figure}

As seen in Figure \ref{fig:z_hist}, archival redshift measurements for BLL 
are dominated by low values ($\tilde{z} = 0.23$). Our 
new spectroscopic redshifts have extended the population to higher $z$, with some objects'
redshifts at $z \gg 1$.  Still, the objects with redshifts remain dominated by low $z$.
In the new spectroscopic redshifts we find $\tilde{z} = 0.33$. We believe there is 
a significant bias to low redshift in both of these samples, as the weak 
low EW emission or absorption features of our typical BLLs with known redshift are easier to 
detect at low $z$.

In Figure \ref{fig:z_hist} we also show two sets of redshift lower limits. For every 
object in our sample, we can derive a host-detection limit ($\tilde{z}=0.41$). These lower 
limits on redshift are still biased low: as evident from Figure \ref{fig:fitter}
we are most sensitive to galaxies at low $z$. Nevertheless, they suggest that 
these objects are {\it not} consistent with the spectroscopic redshifts (the K-S 
test gives probability $ < 10^{-11}$ of consistency with archival redshifts).
The absorption line limits we have for some objects (described in \S 
\ref{sec:intervening}) give further evidence for a population of BLLs at 
high redshift ($z>1$). Together, these results strongly imply that previous 
BLL studies suffered important biases due to shallow samples with large redshift 
incompleteness preventing detection of bright, but high $z$ BLL. 

	The inset shows the spectroscopic redshifts and the redshifts limits for high-peaked
(HBL) and lower-peaked (LBL+IBL) sources classified in 2LAC. We see that the lower-peaked
detections extend to higher z, as might be expected if these sources are more luminous
and have a less dominant synchrotron continuum. However, the open histograms of limits
remind us that both sub-classes still suffer substantial redshift incompleteness, and the
missing redshifts for both sub-classes extend substantially higher than those in hand.
A re-appraisal of the BLL population, properly including
the new redshifts, $z$ constraints and remaining selection biases is needed to 
test whether either subclass is still consistent with negative cosmological evolution.

\section{Black Holes and Host Galaxies}
\label{sec:bh}

The masses of the central black holes provide important insight into the cosmic evolution of various
AGN classes.  These are most easily estimated by the virial technique \citep[cf.,][]{she11}.
In S12, we adopted this method to give mass estimates for the {\it Fermi} FSRQs. 
For BLL, the lack of high S/N broad line measurements precludes such estimates.
However, we have measured a number of host magnitudes in \S \ref{sec:hostfit}; since these 
are ellipticals, this is all `bulge,' and we can apply an $M-L$ relation to estimate the hole 
mass. We follow \citet{gul09}:
\begin{equation}
\log\left(\frac{M}{M_\odot}\right) = (8.95 \pm 0.11) + (1.11 \pm 0.18) \log\left(\frac{L_V}{10^{11}L_{\odot,V}}\right)
\label{eqn:bhmass}
\end{equation}
where $M$ is the black hole mass, and $L_V$ is the luminosity in a $V$ filter 
[$\log(L_V/L_{\odot,V})=0.4(4.83 - M^0_{V,{\rm bulge}})$]. To convert our fit
template spectrum amplitudes to consistent V magnitudes, we integrate our template spectrum 
over the Hubble F555W filter \citep{lau05} as in \S \ref{sec:stdcandle}.

\begin{figure}
\vspace{-6pt}
\hspace*{-19pt}
\epsscale{1.35}
\plotone{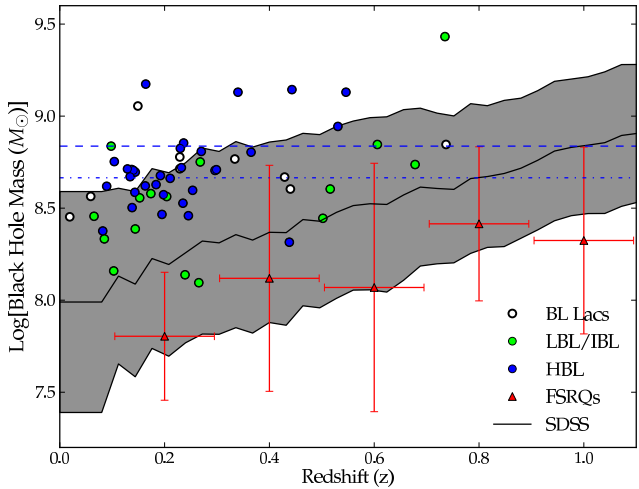}
\vspace{-0pt}
\caption{BLL black hole masses plotted as a function of redshift (circles; blue fill HBL,
green fill LBL+IBL). The gray band 
shows the $1\sigma$ spread of the optically selected SDSS black holes \citep{she11}. 
Red error bars give the $1\sigma$ spread {\it Fermi} FSRQ black hole masses (S12). The estimated
BLL black hole masses are higher than for the two other AGN types.
The upper (lower) blue lines show the black hole mass expected for the standard candle bulge 
at $M_R=-22.9 (-22.5)$.} 
\label{fig:bhmass}
\end{figure}

The masses from  Equation \ref{eqn:bhmass} are plotted as circles in Figure \ref{fig:bhmass}. 
When the sub-class is known (HBL or LBL+IBL) we fill these in with blue or green, respectively.
For comparison, we show the $1\sigma$ spread of virial-estimate BH masses from optically selected 
SDSS quasars from \citet{she11} (gray band) and masses of the {\it Fermi} 
FSRQs in S12 (red points).  Of course, if BLL hosts really are standard candle 
ellipticals, then the $M_{bulge}-M_\bullet$ relation implies constant black hole 
masses. The masses corresponding to the standard $M_R=-22.9$ and our revised
$-22.5$ are shown by dashed lines.

Interestingly, our BLL $M_\bullet$ estimates increase with $z$ much as the optical QSO or FSRQ. 
Of course, we only plot high significance host detections here, and low 
luminosity hosts at high $z$ are increasingly difficult to detect (unless the core 
luminosity decreases). Accordingly,
as for the QSO, we suspect that the bulk of this trend is due to selection effects. In the case of the FSRQ,
S12 argued that the offset to smaller black hole mass was at least partly due to a preferentially polar
view of an equatorially flattened broad line region, with the projection decreasing 
the observed kinematic line width and the average virial mass estimate. Like $\gamma$-ray selected FSRQ,
BLL are Doppler-boosted along our line of sight \citep{urr95}. However since the host flux is 
nearly isotropic, we expect little alignment bias in our $M_\bullet$ estimates. Thus, it is unclear 
whether the BLL offset to larger black hole masses is real or selection dominated.

	In a study of BLL hosts detected in the SDSS \citet{leo11}
found no significant difference between the masses of the central black holes of HBL and LBL.
In figure \ref{fig:bhmass} the HBL masses are however biased upwards with respect to the lower-peak BLL black
hole masses. This is of course just a restatement of the offset in host luminosity seen in
Figure \ref{fig:candlemag}. Unfortunately, we cannot claim that this is a physical difference as
the trend is precisely what would expect from selection bias if HBL have brighter non-thermal cores.

	Ideally we could use these black hole masses to explore the relationship between the BLLs
and the general QSO population. Large black hole masses, if not induced solely by selection bias would imply a late
stage of AGN evolution. The black hole mass is often compared to the source luminosity to characterize the
state of the accretion in Eddington units. However, with the exception of the few BLL for which 
we see broad lines (which seldom have a significantly host detection), the 
observed flux is so dominated by beamed jet emission that quoting the accretion luminosity 
in Eddington units is not feasible. 

\section{Non-thermal Dominance}
\label{sec:ntd}

In S12, we introduced the non-thermal dominance (NTD) as a quantitative measure of how much the 
optical is contaminated by non-thermal synchrotron emission. We here extend that analysis to BLLs.

For most BLL, the dominant `thermal' contribution is the host galaxy, not the 
big blue bump. We therefore set $NTD \equiv F_{\rm core}/F_{\rm host}$ where both fluxes 
are measured at 5100 \AA\ -- the same wavelength as the H$\beta$ continuum measurements for FSRQs.
The wavelength choice is important for BLL NTD measurements, since the host galaxy is much 
redder than the continuum-dominated core; an NTD measurement just above (redward of) the $4000$ \AA\ 
break would typically give values $\gtrsim4\times$ larger. Measurements below the break would, of course, diverge even more.

As noted BLL are highly variable, so that their NTD changes over time. Our BLL can vary by 
over an order of magnitude in the optical (S12, Fig. 3).  In this study, our primary 
spectra for variable objects
were generally drawn from the epochs giving the highest signal to noise on weak emission or 
absorption features. As this favors low core fluxes, our primary epoch biases the results towards low NTD.

\begin{figure}
\vspace{-0pt}
\hspace*{-0pt}
\epsscale{1.2}
\plotone{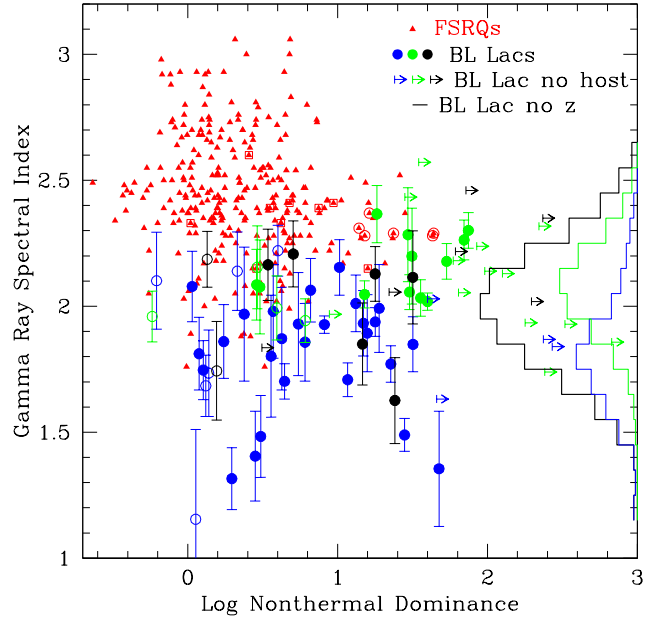}
\vspace{-0pt}
\caption{The NTD plotted against $\gamma$-ray spectral index. The BLL are plotted as dots; 
solid if $L_\gamma > 10^{44}$; empty otherwise. Broad line objects from S12 (typically FSRQs) 
are plotted as red triangles -- circles surround those in a BLL state at the primary
spectrum epoch; squares, those known to be BLL at a different epoch. BLL with known 
$z$, but no thermal (host galaxy or broad emission line) detection are plotted as lower 
limits. BLL without known $z$ are shown by the histogram at right. Blue symbols indicate HBL,
green LBL+IBL.
Except for the NTD
increase for the BLL population relative to the FSRQ, no clear trend with spectral index is seen.}
\label{fig:ntd}
\end{figure}

In Figure \ref{fig:ntd}, we plot the NTD against $\gamma$-ray spectral index. The values for BLL with 
measured galaxy flux are plotted, along with lower limits for BLL with redshift, but no 
significant host detection. The histogram at right shows the spectral index distribution of BLL 
with no redshift detection (and unknown NTD). We expect these BLL to have higher NTD on average, 
as the larger core flux makes redshift determinations more challenging. When known, we indicate
whether the BLL are high-peaked sources (blue) or lower-peaked (green).  The FSRQs from S12 
are plotted as red triangles.

	The most striking trend in this figure is the vertical color separation.
This is the well-known
result that $\gamma$-ray spectral index hardens from FSRQ through LBLs to HBLs \citep{2LAC}.
We also see the defining characteristic of the BLL, increased continuum dominance with respect
to the FSRQ. However, the NTD trend does not appear to continue {\it through} the BLL:
harder spectrum BLL do not in general show increased NTD. Indeed the highest NTD seem to
be associated with LBL/IBL measurements and lower bounds. A plausible interpretation is that
the LBL/IBL are more luminous and hence visible to higher $z$, where we are less likely to detect
a host galaxy.

\section{Conclusions}
\label{sec:conclusion}

We have dramatically increased the redshift completeness of this largest ever $\gamma$-ray 
selected BLL sample; $\sim 44\%$ of these BLL now have spectroscopic redshifts, and 
$\sim 95\%$ have at least a strong lower bound on the redshift. These $z$ constraints show that
the subset with actual spectroscopic redshifts is strongly biased to low $z$. Although
we find that the measured redshifts for low-peaked BLL (LBL+IBL) do extend higher than
those for the HBL with the highest synchrotron peak, our set of lower limits for both
subclasses extend to yet higher $z$ than the spectroscopically solved objects. Thus the
actual redshifts for all BLL are biased low compared to a flux-limited parent population
of $\gamma$-ray BLL. 
This must be taken into account in any study of BLL evolution over cosmic time 
(Ajello et al, in prep).

Many of our redshift limits rely on the common assumption that BLL hosts are standard candles. 
Our effort to re-calibrate the standard luminosity for the {\it Fermi} BLL has resulted
in fainter absolute magnitudes and, hence, more conservative minimum $z$. This implies that
the standard candle
assumption deserves further study, which could be best prosecuted by obtaining more high spatial 
resolution images of BLL with known redshifts. Our study provides a large increase in the
spectroscopic redshifts, a useful precursor to such work. We find limited evidence that
HBL have more luminous hosts than LBL/IBL. Whether this is intrinsic or a selection effect
in the presence of a brighter, harder continuum is not yet clear.

Of course true spectroscopic redshifts are always preferable for uniform population studies.
However, we suspect that much higher completeness will be difficult to attain,
and will likely require novel observational techniques, as significantly more time on $8$\,m 
class telescopes is both expensive and likely to provide only marginally better results.
In the interim, the best hope for progress lies with careful correction for selection effects
in the present sample. Because of the tight correlation of synchrotron peak frequency
with LAT-measured spectral index and because of the $\gamma$-ray index dependence of the
LAT sensitivity, study of the relative population and evolution of different BLL subclasses
will require correction for $\gamma$-ray selection effects as well as possible biases
in the redshift determinations themselves. While we do not attempt such study here, our
improved completeness will help in understanding these selection effects.

We find that BLL black hole mass estimates (at a given redshift) are larger than those for
optically selected quasars or {\it Fermi} FSRQs. Associated with the apparent host luminosity
differences, the present detections suggests that HBL host the largest mass black holes.
We cannot at present tell whether these trends
are a true difference in the black hole populations or luminosity-driven selection effects.

BLLs are, by definition, non-thermally dominated (NTD $>1$). We find the BLL 
population to have significant higher NTD than the {\it Fermi} FSRQs. This claim is conservative,
as the objects without redshift are likely to be at even higher NTD. 
Since BLL have a range of $\gamma$-ray spectral index harder than those of FSRQ, it is 
perhaps surprising that this
trend does not hold {\it within} the BLL class: NTD is not, on average, higher for the
hardest-spectrum BLL. It seems likely that, in this respect at least, BLL are a distinct
population from FSRQ, and not just the hardest spectrum objects on a continuum. In this
case we might expect NTD to be controlled by the precise accident of the jet Doppler boosting.
This accords well with the idea that NTD can vary for an individual source as the jet angle or
effective Doppler factor vary, with relatively little change in the GeV spectrum. Multi-epoch,
multiwavelength studies of the brighter BLL with a range of $\gamma$-ray hardness are needed
to test these ideas.

\acknowledgements

	We thank the referee for a careful reading and comments that helped us add
materially to the paper.

The Hobby-Eberly Telescope (HET) is a joint project of the University of Texas at
Austin, the Pennsylvania State University, Stanford University, Ludwig-Maximilians-Universitaet
Muenchen, and Georg-August-Universitaet Goettingen. The HET is named in honor of its principal
benefactors, William P. Hobby and Robert E. Eberly.
The Marcario Low Resolution Spectrograph is named for Mike Marcario of High Lonesome
Optics, who fabricated several optics for the instrument but died before its completion.
The LRS is a joint project of the Hobby-Eberly Telescope partnership and the Instituto de
Astronomıa de la Universidad Nacional Autonoma de Mexico.

Some of the data presented herein were obtained at the W.M. Keck Observatory, which is operated as a scientific partnership among the California Institute of Technology, the University of California and the National Aeronautics and Space Administration. The Observatory was made possible by the generous financial support of the W.M. Keck Foundation. The authors wish to recognize and acknowledge the very significant cultural role and reverence that the summit of Mauna Kea has always had within the indigenous Hawaiian community.  We are most fortunate to have the opportunity to conduct observations from this mountain.

This work also employs observations obtained at the Southern Astrophysical Research (SOAR) telescope, which is a joint project of the Minist\'{e}rio da Ci\^{e}ncia, Tecnologia, e Inova\c{c}\~{a}o (MCTI) da Rep\'{u}blica Federativa do Brasil, the U.S. National Optical Astronomy Observatory (NOAO), the University of North Carolina at Chapel Hill (UNC), and Michigan State University (MSU).
Additional observations were made with ESO Telescopes at the La Silla Paranal Observatory under programme 
077.B-0056
078.B-0275
079.B-0831
083.B-0460
084.B-0711
087.A-0573. GC acknowledges support from STFC grant ST/H002456/1

	We acknowledge support from NASA grants NNX09AW30G, NXX10AU09G and NAS5-00147. 
A.C.S.R. is also supported under grant AST-0808050.

{\it Facilities:} \facility{Fermi}, \facility{Hale (DBSP)}, \facility{HET}, \facility{KECK:I (LRIS)}, \facility{NTT}, \facility{VLT:Antu (FORS2)}.

\bibliography{references}

\LongTables
\clearpage
\begin{landscape}
\begin{deluxetable*}{ccrcccccccccccccc}
\tablecolumns{17}
\tabletypesize{\tiny}
\tablecaption{BLL Spectral Properties}
\tablehead{
\colhead{2FGL} & \colhead{RA} & \colhead{Dec} & \multicolumn{3}{c}{Name\ \ \ \  \ \ \ $\log F_{\nu,10^{14.7}}$\ \ \ \ \ \ \ \ \ \ $\alpha$\ \ \ \ \ \ } & \colhead{M$_R$} & \colhead{$z$} &  \colhead{ID} & \colhead{$z_{min}$} & \colhead{$z_{min}$} & \colhead{$z_{min}$} & \colhead{$z_{max}$} & \colhead{Type} & \colhead{SED} & \colhead{Tel.} & \colhead{MJD} \\
\colhead{} & \colhead{} & \colhead{} & \multicolumn{3}{c}{$10^{-28}$erg cm$^{-2}$s$^{-1}$Hz$^{-1}$} & \colhead{} & \colhead{} & \colhead{\tablenotemark{a}} & \colhead{\tablenotemark{b}} & \colhead{(-22.5)} & \colhead{(-22.9)} & \colhead{} & \colhead{} & \colhead{} & \colhead{}
}\startdata
J0000.9$-$0748 & 0.325017 & $-$7.774111 & J0001$-$0746 & 1.55$\pm$0.11 & $-$1.399$\pm$0.003 & ... & ... & . & ... & 0.33 & 0.50 & 1.84 & BLL & IBL & P200 & 55064\\
J0007.8+4713 & 1.999862 & 47.202135 & J0007+4712 & 1.33$\pm$0.14 & $-$1.431$\pm$0.430 & ... & 2.1: & S & 1.659 & 0.26 & 0.36 & 2.69 & BLL & LBL & P200 & 55743\\
J0009.0+0632 & 2.267006 & 6.472664 & J0009+0628 & 1.38$\pm$0.11 & $-$1.410$\pm$0.007 & ... & ... & . & ... & 0.58 & 0.76 & 1.65 & BLL & LBL & WMKO & 55040\\
J0009.1+5030 & 2.344750 & 50.508005 & J0009+5030 & 1.51$\pm$0.11 & $-$0.776$\pm$0.010 & ... & ... & . & ... & 0.79 & 0.79 & 1.64 & BLL & ... & WMKO & 55475\\
J0012.9$-$3954 & 3.249804 & $-$39.907184 & J0012$-$3954 & 1.01$\pm$0.11 & $-$1.248$\pm$0.032 & ... & ... & . & ... & 0.52 & 0.64 & 2.52 & BLL & ... & VLT & 54362\\
J0013.8+1907 & 3.485099 & 19.178246 & J0013+1910 & 0.61$\pm$0.11 & $-$2.250$\pm$0.076 & ... & 0.477 & B & ... & 0.41 & 0.54 & 2.17 & BLL & ... & WMKO & 54470\\
J0018.5+2945 & 4.615626 & 29.791743 & J0018+2947 & 1.13$\pm$0.11 & $-$0.701$\pm$0.181 & ... & ... & . & ... & 0.89 & 0.94 & 1.71 & BLL & HBL & WMKO & 55547\\
J0018.8$-$8154 & 4.841042 & $-$81.880833 & J0019$-$8152 & 1.73$\pm$0.12 & $-$0.975$\pm$0.045 & ... & ... & . & ... & 0.24 & 0.34 & 2.32 & BLL & HBL & NTT & 55778\\
J0021.6$-$2551 & 5.385517 & $-$25.846999 & J0021$-$2550 & 1.63$\pm$0.11 & $-$0.689$\pm$0.005 & ... & ... & . & 0.564 & 0.57 & 0.58 & 1.63 & BLL & IBL & WMKO & 55476\\
J0022.2$-$1853 & 5.538158 & $-$18.892486 & J0022$-$1853 & 1.94$\pm$0.11 & $-$1.061$\pm$0.069 & ... & ... & . & 0.774 & 0.13 & 0.15 & 1.64 & BLL & HBL & WMKO & 55476\\
J0022.5+0607 & 5.635264 & 6.134573 & J0022+0608 & 1.86$\pm$0.11 & $-$1.560$\pm$0.009 & ... & ... & . & ... & 0.29 & 0.42 & 1.63 & BLL & LBL & WMKO & 55060\\
J0029.2$-$7043 & 7.170792 & $-$70.754694 & J0028$-$7045 & 1.38$\pm$0.11 & $-$1.139$\pm$0.011 & ... & ... & . & 0.966 & 0.54 & 0.58 & 1.95 & BLL & ... & VLT & 55056\\
J0033.5$-$1921 & 8.393060 & $-$19.359393 & J0033$-$1921 & 1.94$\pm$0.12 & $-$0.852$\pm$0.053 & ... & ... & . & ... & 0.29 & 0.37 & 1.77 & BLL & HBL & P200 & 55567\\
J0035.2+1515 & 8.810913 & 15.250758 & J0035+1515 & 1.93$\pm$0.11 & $-$0.895$\pm$0.003 & ... & ... & . & ... & 0.48 & 0.48 & 1.65 & BLL & HBL & WMKO & 55357\\
J0037.8+1238 & 9.461902 & 12.638514 & J0037+1238 & 2.00$\pm$0.11 & $-$1.163$\pm$0.010 & $-$22.4$\pm$0.1 & 0.089 & G & ... & 0.09 & 0.09 & 1.76 & BLL & HBL & WMKO & 55357
\enddata
\label{tab:props}
\tablecomments{Table \ref{tab:props} is published in its entirety in the electronic 
edition of this journal; A portion is shown here to show its form and content.}
\tablenotetext{a}{Method for $z$ ID---B, broad emission lines; N, narrow emission lines; G, host galaxy features, S, special case -- see \S3.1}
\tablenotetext{b}{Spectroscopic lower limits (i.e.: From intervening absorption systems)}
\end{deluxetable*}
\clearpage
\end{landscape}

\end{document}